\documentclass[
prd
,showpacs,amssymb,superscriptaddress,aps
]{revtex4-2}
\usepackage{graphicx}
\usepackage{color}
\input{epsf}

\usepackage{amsmath,amssymb}
\usepackage{bm}
\usepackage{times}
\usepackage{ulem}

\newcommand{\dalm}{\kern1pt\vbox{\hrule height 0.9pt\hbox{\vrule width 0.9pt
\hskip 2.5pt\vbox{\vskip 5.5pt}\hskip 3pt\vrule width 0.3pt}\hrule height 0.3pt}
\kern1pt}

\begin{document}



\title{Neutron star mass formula with nuclear saturation parameters for asymmetric nuclear matter}

\author{Hajime Sotani}
\email{sotani@yukawa.kyoto-u.ac.jp}
\affiliation{Astrophysical Big Bang Laboratory, RIKEN, Saitama 351-0198, Japan}
\affiliation{Interdisciplinary Theoretical \& Mathematical Science Program (iTHEMS), RIKEN, Saitama 351-0198, Japan}

\author{Shinsuke Ota}
\affiliation{Research Center for Nuclear Physics (RCNP), Osaka University, Ibaragi, Osaka 567-0047, Japan}

\date{\today}

\begin{abstract}
Low-mass neutron stars are directly associated with the nuclear saturation parameters because their central density is definitely low. We have already found a suitable combination of nuclear saturation parameters for expressing the neutron star mass and gravitational redshift, i.e., $\eta\equiv (K_0L^2)^{1/3}$ with the incompressibility for symmetric nuclear matter, $K_0$, and the density-dependent nuclear symmetry energy, $L$. In this study, we newly find another suitable combination given by $\eta_\tau\equiv (-K_\tau L^5)^{1/6}$ with the isospin dependence of incompressibility for asymmetric nuclear matter, $K_\tau$, and derive the empirical relations for the neutron star mass and gravitational redshift as a function of $\eta_\tau$ and the normalized central number density. With these empirical relations, one can evaluate the mass and gravitational redshift of the neutron star, whose central number density is less than threefold the saturation density, within $\sim 10\%$ accuracy, and the radius within a few \% accuracy. In addition, we discuss the neutron star mass and radius constraints from the terrestrial experiments, using the empirical relations, together with those from the astronomical observations. Furthermore, we find a tight correlation between $\eta_\tau$ and $\eta$. With this correlation, we derive the constraint on $K_\tau$ as $-348\le K_\tau\le -237$ MeV, assuming that $L=60\pm 20$ and $K_0=240\pm 20$ MeV.
\end{abstract}

%
\maketitle


\section{Introduction}
\label{sec:I}

Neutron star is a massive remnant left after a supernova explosion, which happens at the last moment of the massive star's life. The density inside the star becomes much higher than the standard nuclear density, $\rho_0=2.7\times 10^{14}$ g/cm$^3$, and the gravitational and magnetic fields inside/around the star are significantly stronger than those observed in our solar system \cite{ST83}. The neutron star mass and radius strongly depend on the equation of state (EOS) for neutron star matter under the $\beta$-equilibrium. The mass of a neutron star model with a higher central density generally becomes larger, even though the EOS is not fixed yet. So, the astronomical observations of neutron stars or their activities tell us the information about the EOS for a relatively higher density region, while the terrestrial experiments tell us that for a lower density region (e.g., Fig. 2 in Ref. \cite{SNN22}). 

In practice, the discovery of $2M_\odot$ neutron stars \cite{D10,A13,C20,F21} has excluded some soft EOSs, with which the expected maximum mass is less than the observed mass. This discovery simultaneously reveals the problem that most of the EOSs with hyperon are too soft to construct the $2M_\odot$ neutron star, i.e., the so-called hyperon puzzle. Meanwhile, the gravitational wave event, GW170817 \cite{gw170817}, tells us the information on the tidal deformability of the neutron star, which leads to the constraint that a $1.4M_\odot$ neutron star radius should be less than 13.6 km \cite{Annala18}. We note that the constraint on neutron star radius may become more stringent in view of the existing multi-messenger observational data \cite{Capano20,Dietrich20}. The light bending due to the strong gravitational field induced by the neutron star is also one of the important phenomena to see the neutron star properties. That is, the pulsar light curve from the rotating neutron star would be modified due to this relativistic effect and one could get the neutron star properties by carefully observing it (e.g., \cite{PFC83,LL95,PG03,PO14,SM18,Sotani20a}). Based on this idea, the Neutron star Interior Composition Explorer (NICER) is now operating on an International Space Station (ISS) and it has already announced the constraint on two neutron stars properties, i.e., PSR J0030+0451 \cite{Riley19,Miller19} and PSR J0740+6620 \cite{Riley21,Miller21}. Furthermore, the direct detection of the gravitational waves from the neutron star in the future may enable us to extract the neutron star properties (e.g., \cite{AK1996,AK1998,STM2001,SH2003,SYMT2011,PA2012,DGKK2013,Sotani2020,SD2021}).

On the other hand, the EOS in a lower density region is also gradually constrained through terrestrial nuclear experiments, but still, there are large uncertainties in EOS parameters (or in neutron star properties) constrained from terrestrial experiments. For instance, the fiducial value of the density-dependent nuclear symmetry energy $L$ is $L\simeq 60\pm 20$ MeV \cite{Vinas14,BALi19}, while the constraints of $L$ obtained recently seem to be significantly larger than the fiducial value  \cite{SPIRIT,PREXII}. This is because one has to usually transform the experimental constraint to the EOS parameters, even if the information determined via experiments is associated with some aspects of nuclear EOS. Then, one can eventually discuss the neutron star mass and radius as a solution of the Tolman-Oppenheimer-Volkoff (TOV) equation. Anyway, the terrestrial experiments are definitely crucial for understanding the neutron star EOS as well as the astronomical observation of neutron stars. 

In order to discuss the neutron star mass and radius directly with the EOS parameters constrained somehow from the experiments, it may be useful to derive a kind of empirical formula expressing the neutron star properties as a function of EOS parameters if it exists. In practice, we have already found the empirical relations expressing the mass ($M$) and gravitational redshift ($z$) of a low-mass neutron star as a function of the suitable combination of the nuclear saturation parameters, $\eta$, and the central density of neutron star \cite{SIOO14}, with which the neutron star mass and radius expected from the terrestrial experiments can be discussed as in Ref. \cite{SNN22}. On the other hand, in this study, we consider deriving another type of empirical relations for the low-mass neutron stars, focusing on the nuclear saturation parameters for asymmetric nuclear matter, and then discuss the impact of the uncertainties in the nuclear saturation parameters on the neutron star mass and radius.

This manuscript is organized as follows. In Sec. \ref{sec:EOS}, we briefly mention the EOSs considered in this study and the saturation parameters in nuclear matter. In Sec. \ref{sec:NS}, we examine the neutron star models and derive the empirical formulas for the neutron star mass and its gravitational redshift as a function of a suitable combination of the nuclear saturation parameters we derived in this study. Using the newly derived empirical formulas, in Sec. \ref{sec:MR}, we discuss the neutron star mass and radius together with the constraints from the astronomical and experimental observations. Finally, in Sec. \ref{sec:Conclusion}, we conclude this study. Unless otherwise mentioned, we adopt geometric units in the following, $c=G=1$, where $c$ and $G$ denote the speed of light and the gravitational constant, respectively.

\begin{table*}
\caption{EOS parameters adopted in this study, $K_0$, $n_0$, $L$, $Q$, $K_{sym}$, $Q_{sym}$, $K_\tau$, and $Q_\tau$ are listed, while $\eta$ and $\eta_\tau$ are specific combinations with them given by $\eta = \left(K_0 L^2\right)^{1/3}$ and $\eta_\tau = \left(-{K_\tau}L^5\right)^{1/6}$. 
} 
\label{tab:EOS2}
\begin {center}
\begin{tabular}{c|cccccccc|cc}
\hline\hline
EOS &  $K_0$ & $n_0$ & $L$ & $Q$  & $K_{sym}$ & $Q_{sym}$ & $K_\tau$ & $Q_\tau$ & $\eta$ & $\eta_\tau$     \\
          & (MeV)& (fm$^{-3}$) & (MeV) & (MeV) &  (MeV)  &  (MeV) &  (MeV)  & (MeV) & (MeV) & (MeV) \\
\hline
OI-EOSs &  200 & 0.165 & 35.6  & -759 &  -142   & 801  & -221 &  2017 & 63.3  & 48.2    \\  
               &         & 0.165 & 67.8  & -761 &  -27.6  & 589  & -176 &  2909 & 97.2  &  79.5   \\  
               &  220  & 0.161 & 40.2 & -720 & -144    & 731  & -254 & 1915 & 70.9  & 54.7    \\  
               &          & 0.161 & 77.6 & -722 & -9.83   & 486  & -221 & 2779 & 110   &  92.4   \\  
               &  240  & 0.159 & 45.0 & -663 & -146    & 642  & -291 & 1760 & 78.6 &  61.4    \\  
               &          & 0.158 & 88.2 & -664 & 10.5    & 363  & -274 & 2559 & 123  &  107     \\  
               &  260  & 0.156 & 49.8 & -589 & -146    & 535  & -333 & 1551 & 86.4 &  68.4    \\  
               &          & 0.155 &  99.2 & -590 &  32.6  & 219  & -338 & 2246 & 137 &   122     \\  
               &  280  & 0.154 & 54.9  & -496 & -146   & 410  & -378 & 1285 & 94.5 & 75.7     \\  
               &          & 0.153 & 111   & -498 & 57.4    & 54.4 & -412 & 1834 & 151  & 138      \\  
               &  300  & 0.152 & 60.0 & -386 & -146    & 266  & -429 & 962 &  103 & 83.3     \\  
               &          & 0.151 & 124  & -387 & 86.1    & -133 & -499 & 1310 & 167  &  157     \\  
 KDE0v  & 229   & 0.161 & 45.2 &  -373 & -145   & 523   & -342 & 1187 &  77.6 & 63.4    \\  
 KDE0v1 & 228 & 0.165  & 54.7  & -385 & -127   &  484  & -363 & 1317 & 88.0 &  75.0    \\  
 SLy2      & 230   & 0.161 & 47.5 & -364 & -115   &  507  & -325 & 1183 &  80.3 & 65.4    \\  
 SLy4      &  230  & 0.160 & 45.9 & -363 & -120   &  522  & -323 & 1175 &  78.7 & 63.7    \\  
 SLy9      &  230  & 0.151 & 54.9 & -350 &  -81.4 &  462  & -327 & 1215 &  88.4 & 73.9    \\  
 SKa       &  263  & 0.155 & 74.6 & -300 &  -78.5  & 175  & -441 &  940  &  114  &  100      \\  
 SkI3       & 258 & 0.158 & 101   & -304  & 73.0   &  212  & -412 & 1276 &  138  &  127      \\  
 SkMp     & 231 & 0.157 & 70.3  & -338  & -49.8  &  159  & -369 & 1086 &  105 &  92.7     \\  
\hline \hline
\end{tabular}
\end {center}
\end{table*}

\section{EOS for neutron star matter}
\label{sec:EOS}

To construct the neutron star models theoretically, one has to prepare the EOS for neutron star matter. We note that, in order to discuss the neutron star properties with the EOS parameters, one has to adopt the unified EOS. That is, the EOS describing the neutron star core is constructed with the same nuclear model as in the EOS for the neutron star crust. 
We note that, if one constructs the neutron star model with the EOS (unlike the unified EOS), which is assembled by connecting the EOS for the core region to the different EOS for the crust region at an appropriate transition density, the radius of a neutron star whose central density is around the transition density strongly depends on the selection of the transition density (e.g., Ref. \cite{Fortin16}). That is, unless the unified EOSs are adopted, one can not discuss the empirical relations, expressing the neutron star models, as we will derive in Sec.~\ref{sec:NS}.
In this study, to systematically see the EOS dependence of the neutron star properties, we particularly adopt the phenomenological EOS proposed by Oyamatsu and Iida \cite{OI03,OI07} (hereafter referred to as OI-EOS) and the EOSs with the Skyrme-type interaction listed in Table \ref{tab:EOS2}.

OI-EOSs are based on the Pad\'{e}-type potential energies and constructed in such a way that the nucleus models with a simplified version of the extended Thomas-Fermi theory should become consistent with the empirical masses and radii of stable nuclei. We note that most of the OI-EOSs adopted here may not fulfill either constraint of $2M_\odot$ observation or the $1.4M_\odot$ neutron star radius constrained from the GW170817, because the OI-EOSs are constructed by focusing on the behavior around the saturation point. Nevertheless, since we discuss the neutron star properties constructed with the central density, which is not so high, and since the OI-EOSs are definitely suitable for systematical study, we adopt the OI-EOSs in this study. On the other hand, we also consider the EOSs with various Skyrme-type interactions, such as KDE0v, KDE0v1 \cite{KDE0v}, SLy2, SLy4, SLy9 \cite{SLy4,SLy9}, SKa \cite{SKa}, SkI3 \cite{SkI3}, and SkMp \cite{SkMp}.
We note that the non-relativistic EOSs may break the causality in a high density region, where the resultant neutron star models are not realistic. But, in this study, we focus only on the density region less than threefold the saturation density, where all the EOSs adopted in this study do not break the causality.  

Even though the EOS generally depends on the nuclear model, compositions, and interaction, the bulk energy per nucleon for the uniform nuclear matter at zero temperature can anyhow be expressed as a function of the baryon number density, $n_{\rm b}$, and an asymmetry parameter, $\alpha$, such as
\begin{equation}
  \frac{E}{A} = w_s(n_{\rm b}) + \alpha^2 S(n_{\rm b}) + {\cal O}(\alpha^3), \label{eq:E/A}
\end{equation}
where $n_b$ and $\alpha$ are given by $n_{\rm b}=n_n+n_p$ and $\alpha=(n_n-n_p)/n_{\rm b}$ with the neutron number density, $n_n$, and the proton number density, $n_p$; $w_s$ corresponds to the energy per nucleon of symmetric nuclear matter ($\alpha=0$); and $S$ denotes the density-dependent symmetry energy. Additionally, $w_s$ and $S$ can be expanded around the saturation density, $n_0$, of the symmetric nuclear matter as a function of $u=(n_{\rm b}-n_0)/(3n_0)$;
\begin{gather}
  w_s(n_{\rm b}) = w_0 + \frac{K_0}{2}u^2 + \frac{Q}{6}u^3 + {\cal O}(u^4), \label{eq:ws} \\
  S(n_{\rm b}) = S_0 + Lu + \frac{K_{sym}}{2}u^2 + \frac{Q_{sym}}{6}u^3 + {\cal O}(u^4). \label{eq:S}
\end{gather}
The coefficients in this expansion correspond to the nuclear saturation parameters, and each EOS has its own set of nuclear saturation parameters. Among the nuclear saturation parameters, $n_0$, $w_0$, and $S_0$ are especially well-constrained from the terrestrial experiments, i.e., $n_0\approx 0.15-0.16$  fm$^{-3}$, $w_0 \approx -15.8$ MeV \cite{OHKT17}, and $S_0 \approx 31.6 \pm 2.7$ MeV \cite{BALi19}. The constraint on $K_0$ and $L$ is relatively more difficult. This is because one needs to obtain the nuclear data in a wide density range to constrain $K_0$ and $L$, which are a kind of density derivative. The current fiducial values of $K_0$ and $L$ are $K_0=240\pm 20$ MeV \cite{Sholomo} and $L=60\pm 20$ MeV \cite{Vinas14,BALi19}. 
Meanwhile, since the determination of the values of $K_0$ and $L$ are strongly model-dependent, one may consider $K_0 = 200-315$ MeV \cite{SSM14,Wang18} and $L=20-145$ MeV \cite{SNN22} as their conservative values, which cover almost all predictions from various experiments.
The experimental constraints on the saturation parameters for higher order terms, such as $Q$, $K_{sym}$, and $Q_{sym}$, are almost disorganized, but they are theoretically evaluated as $-800 \le Q \le 400$, $-400 \le K_{sym} \le 100$, and $-200 \le Q_{sym} \le 800$ MeV \cite{BALi19}.

On the other hand, considering $E/A$ as  the energy per particle of infinite asymmetric nuclear matter, one can also expand it around an isospin dependent saturation density, $\tilde{n}_0(\alpha)\simeq n_0\left[1-3(L/K_0)\alpha^2\right]$ \cite{Vidana09}, such as
\begin{equation}
  \frac{E}{A} = \tilde{w}_0(\tilde{n}_0) + \frac{\tilde{K}_0(\tilde{n}_0)}{2} \tilde{u}^2
     + \frac{\tilde{Q}(\tilde{n}_0)}{6}\tilde{u}^3 + {\cal O}(\tilde{u}^4), \label{eq:E_ANM}
\end{equation}
where $\tilde{u}$ is defined as $\tilde{u}=(n_{\rm b}-\tilde{n}_0)/(3\tilde{n}_0)$ and the coefficients in this expansion are related to the saturation parameters in Eqs. (\ref{eq:ws}) and (\ref{eq:S}) through
\begin{gather}
  \tilde{w}_0(\tilde{n}_0) = w_0 + S_0\alpha^2 + {\cal O}(\alpha^3),  \\
  \tilde{K}_0(\tilde{n}_0) = K_0 +  K_\tau\alpha^2 + {\cal O}(\alpha^3), \\
  \tilde{Q}_0(\tilde{n}_0) = Q + Q_\tau\alpha^2 + {\cal O}(\alpha^3).  
\end{gather}
Here, $K_\tau$ and $Q_\tau$ are respectively the isospin dependence of incompressibility and skewness coefficient at the saturation density, $\tilde{n}_0(\alpha)$, given by
\begin{gather}
   K_\tau = K_{sym} - 6L - \frac{Q}{K_0}L, \label{eq:Ktau} \\
   Q_\tau = Q_{sym} -\frac{9Q}{K_0}L.
\end{gather}

\begin{figure}[tbp]
\begin{center}
\includegraphics[scale=0.5]{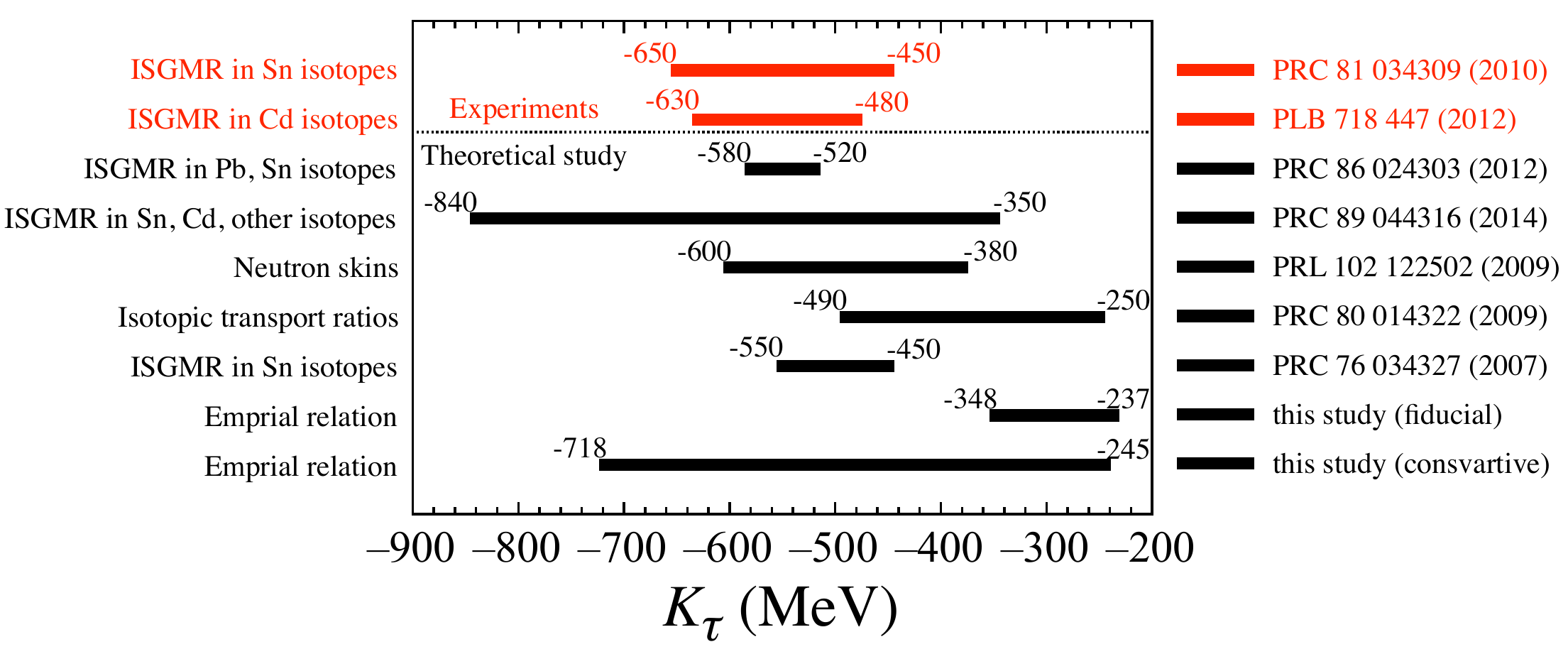}
\end{center}
\caption{
Experimental and theoretical constraints on $K_\tau$ obtained so far. The constraint of $-348\le K_\tau\le -237$ MeV ($-718\le K_\tau\le-245$ MeV) is obtained from the empirical relation given by Eq. (\ref{eq:etat_fit}), assuming that $L=60\pm 20$ MeV \cite{Vinas14,BALi19} and $K_0=240\pm 20$ MeV \cite{Sholomo} ($L=20-145$ MeV \cite{SNN22} and $K_0=200-315$ MeV \cite{SSM14,Wang18}).
}
\label{fig:Ktau}
\end{figure}

Compared to the constraints on $K_0$ and $L$, the constraints on $K_\tau$ are still very poor  \cite{GC18,RP18}. 
The constraints obtained from the experiments are $K_\tau=-550\pm 100$ MeV \cite{Li10} from the analysis of the Isoscalar Giant Monopole Resonance (ISGMR) in Sn isotopes and $K_\tau=-555\pm 75$ MeV from the the analysis of the ISGMR in Cd isotopes \cite{Patel12}, which are performed at the Research Center for Nuclear Physics (RCNP), Osaka University, Japan. On the other hand, using the experimental data in RCNP together with those in the Texas A\&M University (TAMU), the theoretical constraints are also obtained:  $K_\tau=-550\pm 30$ MeV from the analysis of the ISGMR in Pb and Sn isotopes \cite{Vesely12}; $-840<K_\tau<-350$ MeV from the reanalysis of ISGMR in Sn, Cd, and data on $58\le A\le 208$ nuclei   \cite{Stone14}; $K_\tau=-500_{-100}^{+120}$ MeV from the analysis of the neutron-skin data for different anti-protonic atoms \cite{Centelles09}; $K_\tau=-370\pm 120$ MeV from the analysis of the isotopic transport ratios \cite{Chen09}; and $K_\tau=-500\pm 50$ MeV from the analysis of the ISGMR in Sn isotopes \cite{Sagawa07}. In Fig. \ref{fig:Ktau} we show the experimental and theoretical constraints on $K_\tau$ obtained so far. In this study we particuraly adopt the result obtained in Ref. \cite{Li10}, i.e., $K_\tau=-550\pm 100$ MeV, as a fiducial value of $K_\tau$.

\begin{figure}[tbp]
\begin{center}
\includegraphics[scale=0.5]{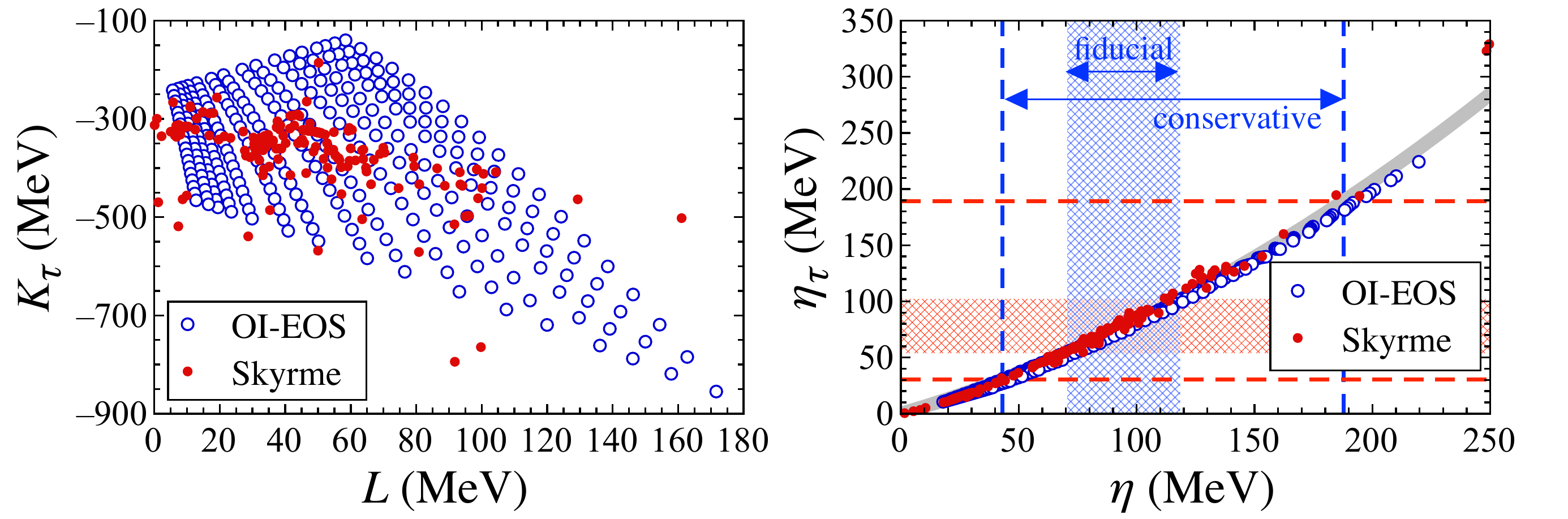}
\end{center}
\caption{
The left panel is the relation between $L$ and $K_\tau$  for the OI-EOSs (open circles) and the Skyrme type EOS models (filled circles). The right panel is the relation between $\eta$ and $\eta_\tau$. The thick-solid line denote the fitting formula given by Eq.~(\ref{eq:etat_fit}). With this empirical relation together with the fiducial value of $\eta$, i.e., $70.6\le\eta\le118.5$ MeV (shaded region), the expected value of $\eta_\tau$ is in the range of $53.8\le\eta_\tau\le102.2$ MeV (shaded region). In a similar way, the value of $\eta_\tau$ is expected in the range of $30.4\le\eta_\tau\le 189.3$ MeV (the region between two horizontal dashed lines) with the conservative value of $\eta$, i.e., $43.1\le\eta\le 187.8$ MeV (the region between two vertical dashed lines).
}
\label{fig:eta-etat}
\end{figure}

Since the neutron star properties are determined by solving the balance equations, i.e., the TOV equation,  together with the EOS depending on several physical inputs, it may generally be difficult to predict the neutron star properties as a function of the EOS parameters. Nevertheless, we could find a suitable combination of the nuclear saturation parameters, $\eta$, for expressing the low-mass neutron star \cite{SIOO14}. Moreover, as mentioned in the following section, we newly find another suitable combination of the saturation parameters, $\eta_\tau$, for expressing the neutron stars. These suitable combinations, $\eta$ and $\eta_\tau$, are given by 
\begin{gather}
  \eta = \left(K_0 L^2\right)^{1/3}, \label{eq:eta} \\
  \eta_\tau = \left(-K_\tau L^5\right)^{1/6}. \label{eq:etatau}
\end{gather}
We note that the finding of $\eta$ and $\eta_\tau$ is just through trial and error, and unfortunately we could not understand the physics behind these quantities.
Since one can express the neutron star properties with $\eta$ or $\eta_\tau$, it is natural to expect the existence of a tight correlation between $\eta$ and $\eta_\tau$. In fact, there is no correlation between $K_\tau$ and $L$ as shown in the left panel of Fig. \ref{fig:eta-etat}, while we find the tight correlation between $\eta$ and $\eta_\tau$ as shown in the right panel of Fig. \ref{fig:eta-etat}, adopting the 304 models for OI-EOSs and 240 models for the Skyrme-type EOSs listed in Ref. \cite{Dutra12}. We note that, for plotting Fig. \ref{fig:eta-etat}, we omit some of the 240 models in Ref. \cite{Dutra12} with the negative value of $L$. From this figure, we can derive the empirical relation, such as 
\begin{equation}
  \eta_\tau = 61.439 \eta_{100} + 20.965 \eta_{100}^2, \label{eq:etat_fit}
\end{equation}
where $\eta_{100}$ is defined by $\eta_{100}\equiv \eta/(100\ {\rm MeV})$. From this empirical relation together with the fiducial value of $\eta$, i.e., $70.6\le\eta\le118.5$ MeV ($L=60\pm 20$ MeV \cite{Vinas14,BALi19} and $K_0=240\pm 20$ MeV \cite{Sholomo}), one can expect $53.8\le\eta_\tau\le102.2$ MeV, which leads to $-348\le K_\tau\le -237$ MeV. 
In a similar way, using $30.4\le\eta_\tau \le 189.3$ MeV estimated with the conservative value of $\eta$, i.e., $43.1\le\eta\le 187.8$ MeV ($L=20-145$ MeV \cite{SNN22} and $K_0=200-315$ MeV \cite{SSM14,Wang18}), one can get the constraint on $K_\tau$ as $-718\le K_\tau\le -245$ MeV. 
These constraints on $K_\tau$ are also shown in Fig. \ref{fig:Ktau}. 
From Fig.~\ref{fig:Ktau}, one can observe that the value of $K_\tau$ constrained with the conservative values of $K_0$ and $L$ through Eq. (\ref{eq:etat_fit}) is more or less similar to the conservative value of $K_\tau$.

\section{Neutron star mass formula}
\label{sec:NS}

In the previous studies, we could derive the empirical formula expressing the neutron star mass and gravitational redshift as a function of $\eta$ and $u_c=\rho_c/\rho_0$ (or $n_c/n_0$), where $\rho_c$ and $\rho_0$ are respectively the central density of neutron star and the saturation density, while $n_c$ is the baryon number density at the stellar center \cite{SIOO14,ST22}. With $\eta$, one can also discuss the rotational properties in low-mass neutron stars \cite{SSB16} and the possible maximum mass of neutron stars \cite{Sotani17,SK17}. 

\begin{figure}[tbp]
\begin{center}
\includegraphics[scale=0.5]{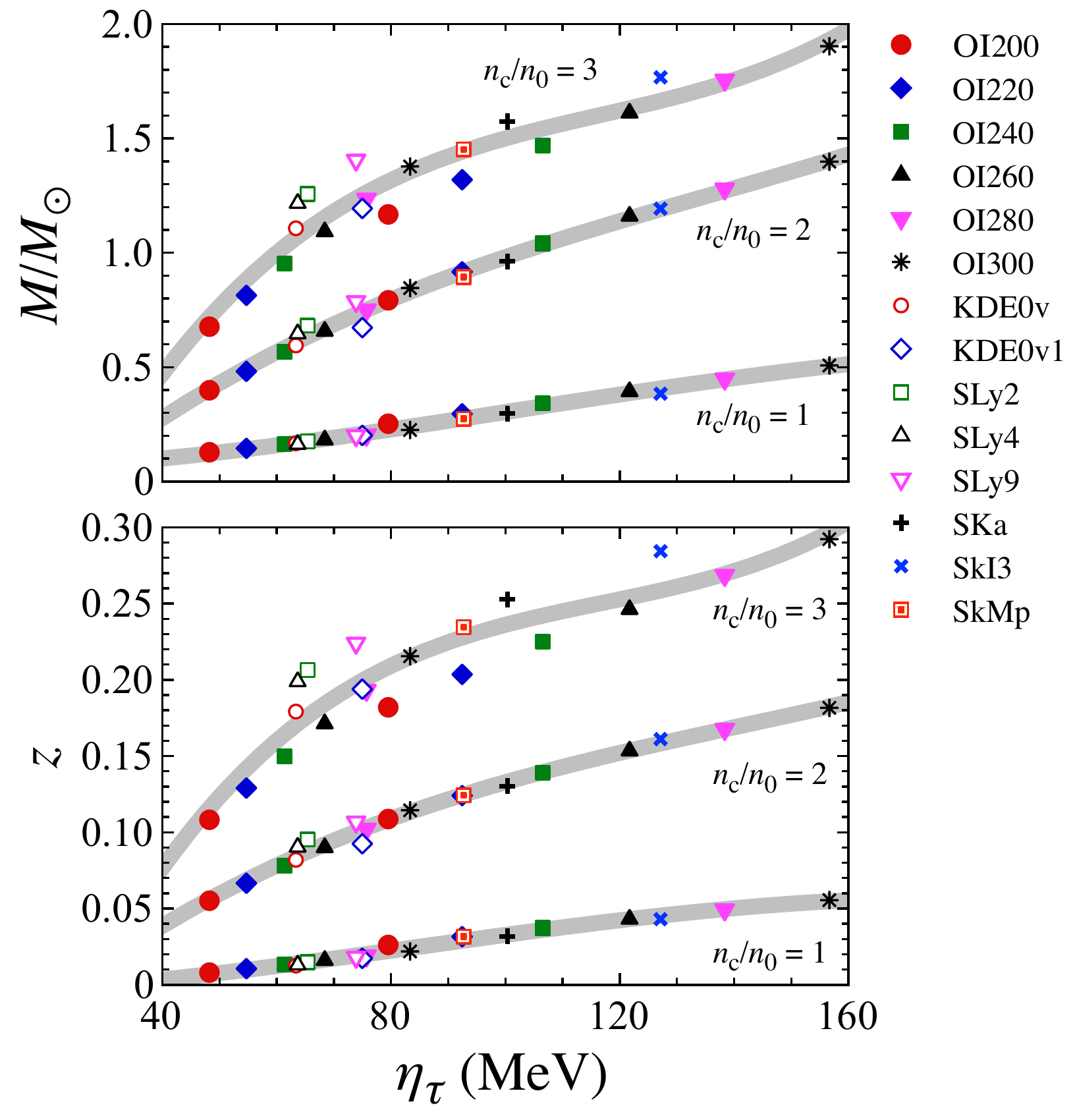}
\end{center}
\caption{
The mass (top panel) and gravitational redshift (bottom panel) for the neutron star models with $n_c/n_0=1$, 2, and 3, constructed with various EOSs, are plotted as a function of $\eta_\tau$. The thick-solid lines are fitting lines given by Eqs. (\ref{eq:Meta}) and (\ref{eq:zeta}).
}
\label{fig:Mzetat}
\end{figure}

Now, we try to derive another type of empirical relation for the mass and gravitational redshift of the neutron star in a similar way considered in Ref. \cite{SIOO14,ST22}. The neutron star mass is determined by integrating the TOV equation, assuming the central density, where the resultant mass also depends on the EOS parameters. As a result of trial and error by hand, we find a suitable combination of $K_\tau$ and $L$, i.e., $\eta_\tau$ given by Eq. (\ref{eq:etatau}), with which the neutron star mass with a fixed central number density can be characterized well, as shown in the top panel of Fig. \ref{fig:Mzetat}. In the same way, we find that the gravitational redshift, $z$, given by $z=(1-2GM/Rc^2)^{-1/2}-1$, for a fixed central number density is also well expressed as a function of $\eta_\tau$, as shown in the bottom panel of Fig. \ref{fig:Mzetat}. From this result, we can derive the fitting formulae for the neutron star mass and gravitational redshift for each central number density as a function of $\eta_\tau$, such as 
\begin{gather}
  M/M_\odot = a_{0}^m + a_{1}^m\eta_{\tau,100} + a_{2}^m\eta_{\tau,100}^2 + a_{3}^m\eta_{\tau,100}^3,
      \label{eq:Meta} \\
  z = a_{0}^z + a_{1}^z\eta_{\tau,100} + a_{2}^z\eta_{\tau,100}^2 + a_{3}^z\eta_{\tau,100}^3,
      \label{eq:zeta} 
\end{gather}
where $\eta_{\tau,100}$ is defined by $\eta_{\tau,100}\equiv \eta_\tau/(100\ {\rm MeV})$, while $a_{i}^m$ and $a_{i}^z$ for $i=0-3$ are coefficients depending on $n_c/n_0$. 

\begin{figure}[tbp]
\begin{center}
\includegraphics[scale=0.5]{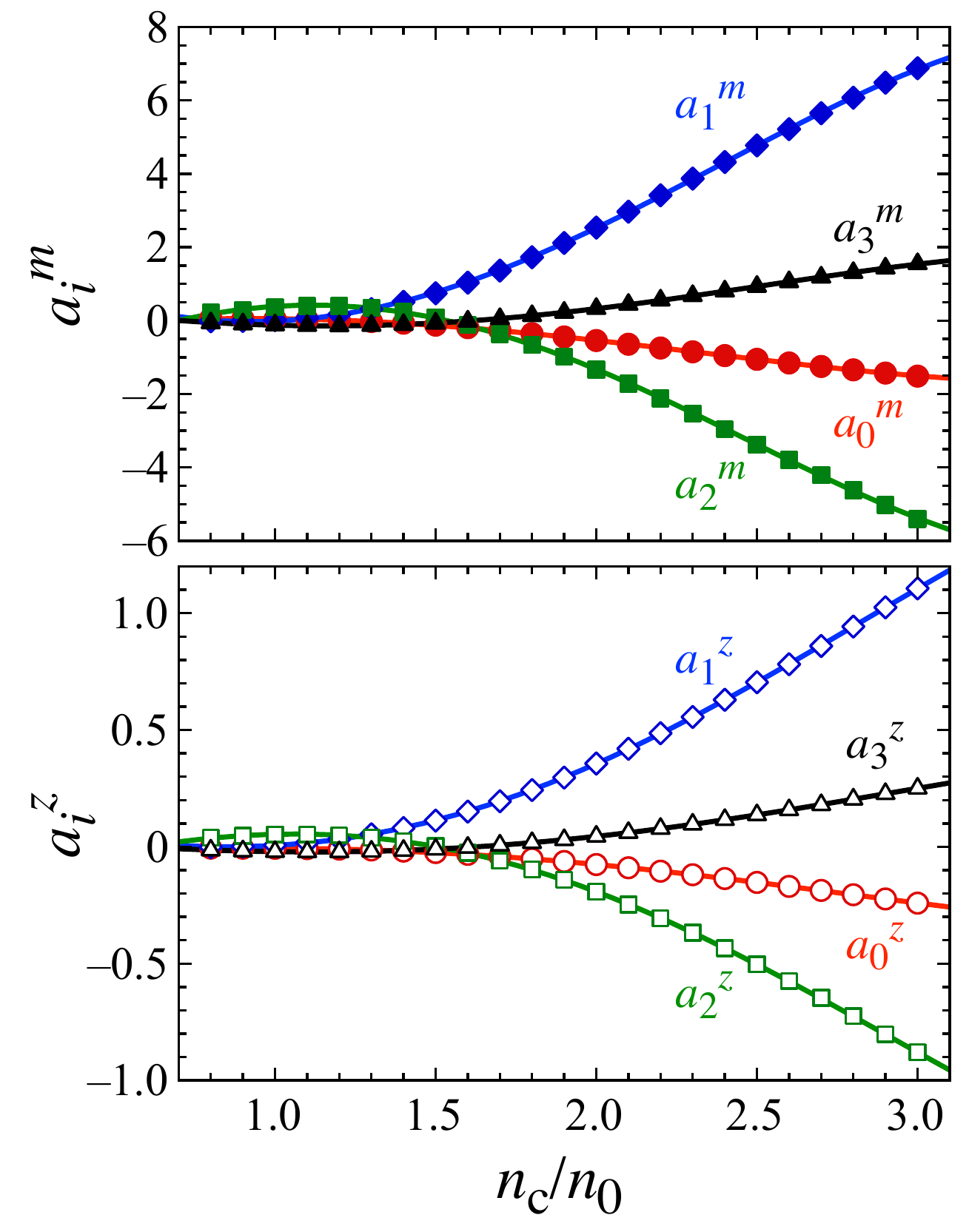}
\end{center}
\caption{
The coefficients in the fitting formula (\ref{eq:Meta}) and (\ref{eq:zeta}) are shown as a function of $n_c/n_0$, where the top and bottom panels correspond to the results for $a_{i}^m$ and $a_{i}^z$, respectively. In the both panels, the solid lines denote the fitting of $a_{i}^m$ and $a_{i}^z$ given by Eqs. (\ref{eq:ami}) and (\ref{eq:azi}).
}
\label{fig:amzi}
\end{figure}

Moreover, the coefficients in Eqs. (\ref{eq:Meta}) and (\ref{eq:zeta}), i.e., $a_{i}^m$ and $a_{i}^z$, are calculated by varying $n_c/n_0$, which are shown in Fig. \ref{fig:amzi}. From this result, we can derive the fitting formulae for $a_{i}^m$ and $a_{i}^z$ as a function of $n_c/n_0$, such as
\begin{gather}
  a_{i}^m = \sum_{j=0}^4 a_{ij}^m(n_c/n_0)^j,  \label{eq:ami} \\
  a_{i}^z =  \sum_{j=0}^4 a_{ij}^z(n_c/n_0)^j.  \label{eq:azi}
\end{gather}
The concrete values of the coefficients in these equations, $a_{ij}^m$ and $a_{ij}^z$, are listed in Table \ref{tab:coefficients}. At last, we can derive the empirical relations for neutron star mass and gravitational redshift as a function of $\eta_\tau$ and $n_c/n_0$, given by Eqs. (\ref{eq:Meta}) -- (\ref{eq:azi}).

\begin{table}
\caption{Values of $a_{ij}^m$ and $a_{ij}^z$ in Eqs. (\ref{eq:ami}) and (\ref{eq:azi}).} 
\label{tab:coefficients}
\begin {center}
\begin{tabular}{c|ccccc}
\hline\hline
 $j$  &  0 & 1 & 2  & 3  & 4   \\
\hline
 $a_{0j}^m$ &  $-0.4498$  & $1.0714$  & $-0.5067$  & $-0.09871$  &  $0.03648$  \\
 $a_{1j}^m$ &  $2.2712$  & $-4.9277$  & $2.4996$  & $0.2726$  &  $-0.1296$  \\
 $a_{2j}^m$ &  $-2.4238$  & $4.6837$  & $-1.2923$  & $-0.7641$  &  $0.1883$  \\
 $a_{3j}^m$ &  $0.6795$  & $-1.1991$  & $0.1422$  & $0.3175$  &  $-0.06662$  \\
 \hline
 $a_{0j}^z$  &  $-0.03506$  & $0.04173$  & $0.01848$  & $-0.03742$ &  $0.006334$  \\
 $a_{1j}^z$  &  $0.1561$  & $-0.3211$  & $0.09294$  & $0.09648$ &  $-0.01888$  \\
 $a_{2j}^z$  &  $-0.1884$  & $0.3743$  & $-0.03323$  & $-0.1200$ &  $0.02129$  \\
 $a_{3j}^z$  &  $0.05780$  & $-0.1110$  & $-0.001509$  & $0.04061$ &  $-0.006882$  \\
\hline \hline
\end{tabular}
\end {center}
\end{table}

\begin{figure}[tbp]
\begin{center}
\includegraphics[scale=0.5]{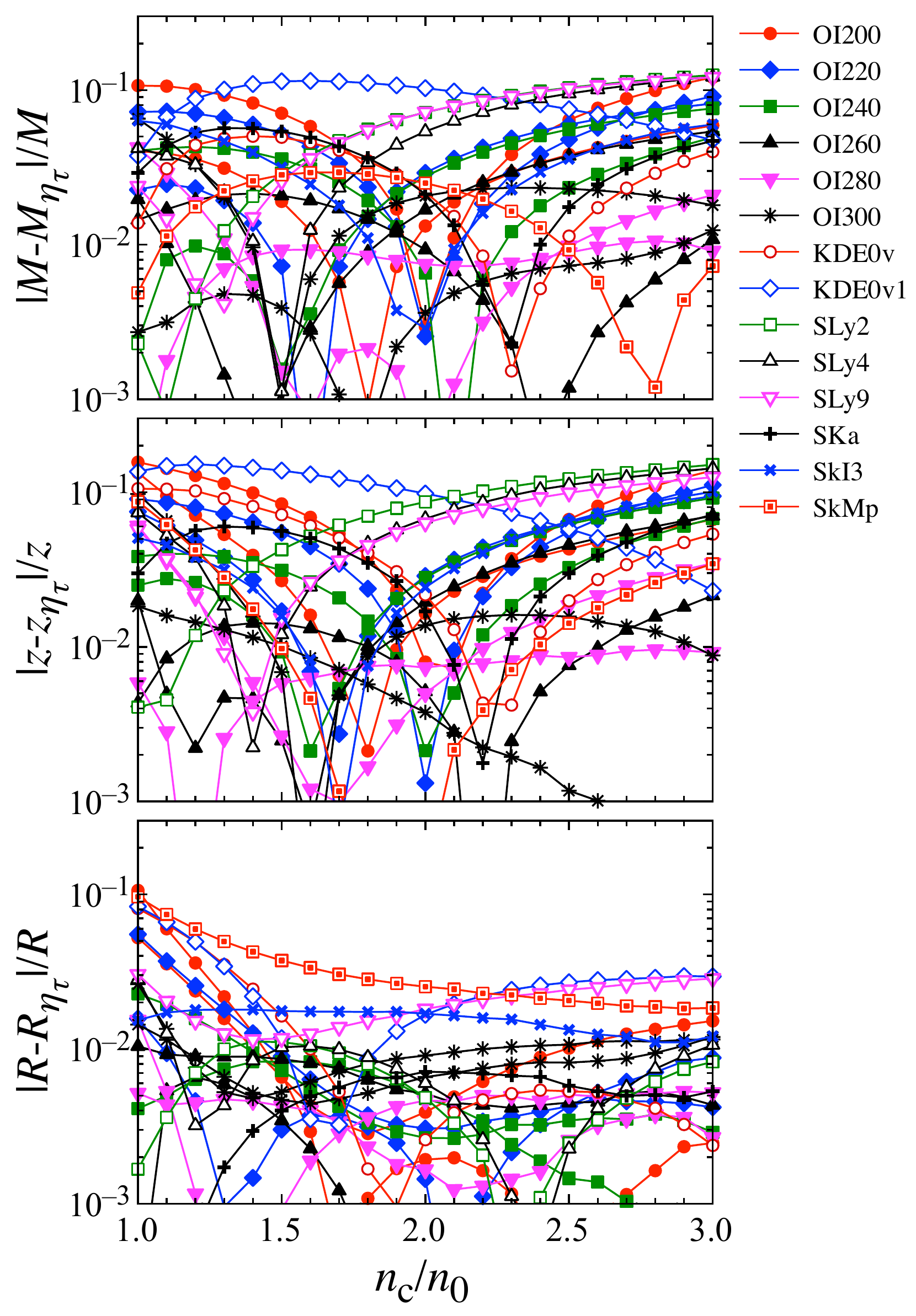}
\end{center}
\caption{
The relative deviation of neutron star mass (top panel) and gravitational redshift (middle panel) estimated with the empirical formulas from those determined with the specific EOSs is shown as a function of $n_c/n_0$. The bottom panel is the relative deviation of the neutron star radius estimated with the empirical formulas for the stellar mass and gravitational redshift from that of the TOV solution. 
}
\label{fig:dMzR}
\end{figure}

Here, we check how well the empirical relations derived in this study work. In Fig. \ref{fig:dMzR} we show the relative deviation of the neutron star mass (top panel), gravitational redshift (middle panel), and radius (bottom panel) estimated with the empirical relations from those for a TOV solution with various EOSs. From this figure, one can observe that the neutron star mass and gravitational redshift are estimated with the empirical relations within $\sim 10\%$ accuracy, while the radius for neutron star models with $n_c/n_0\ge 1.7$ is estimated within a few \% accuracy.

\section{Neutron star mass and radius relation}
\label{sec:MR}

To simultaneously discuss the EOS parameters constrained from the terrestrial experiments together with the EOSs for a higher density region constrained from the astronomical observations, the neutron star mass and radius must be reasonable. In this section, first, we briefly mention the constraints on the neutron star mass and radius obtained from several astronomical observations. Then, we discuss the impact of the EOS parameters constrained from the terrestrial experiments on the neutron star mass and radius by using the empirical formulae, Eqs. (\ref{eq:Meta}) -- (\ref{eq:azi}), derived in this study.

\subsection{Constraints from astronomical observations}
\label{sec:MR-astro}

\begin{figure}[tbp]
\begin{center}
\includegraphics[scale=0.6]{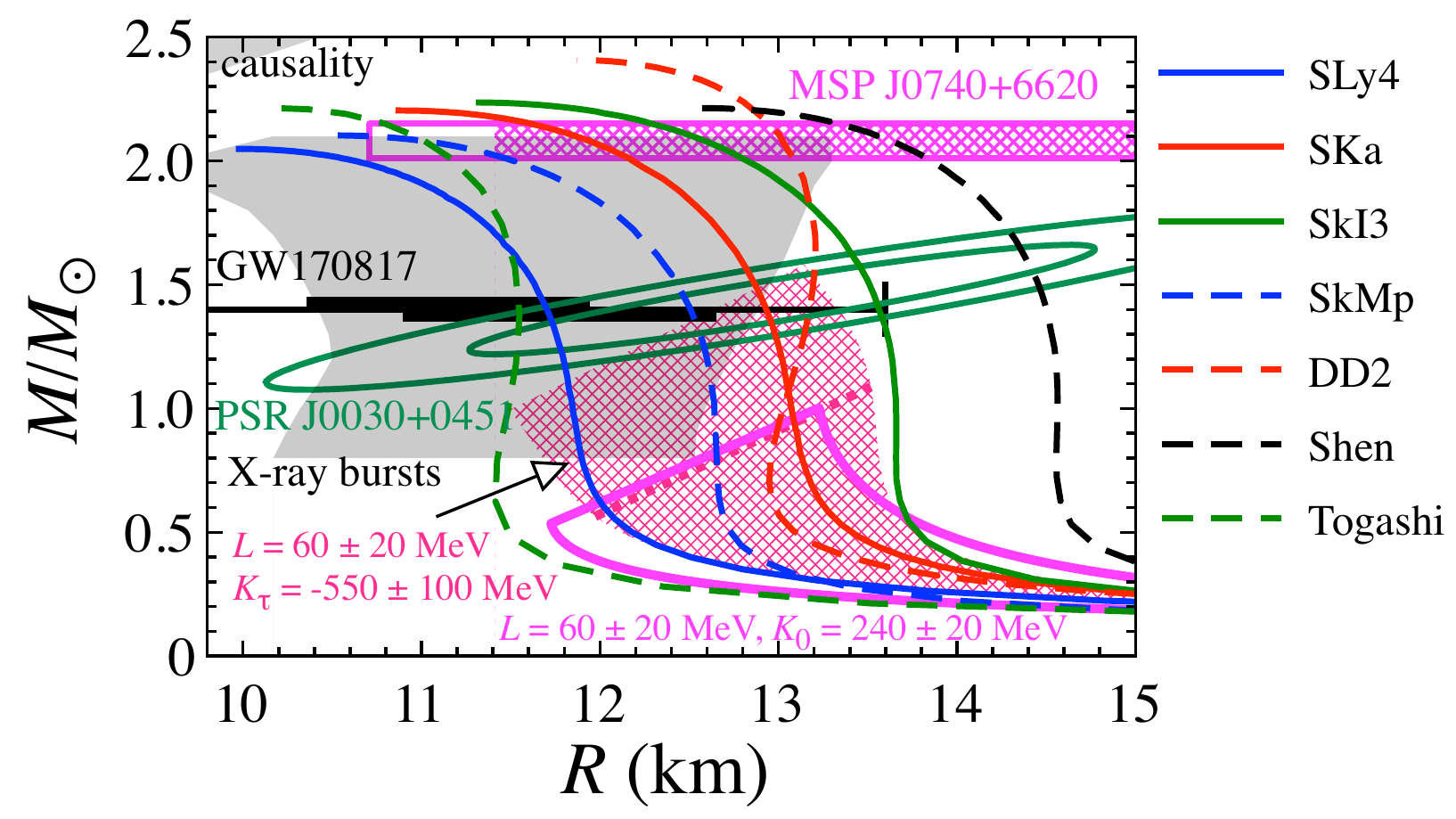}
\end{center}
\caption{
The expected region in the relation between neutron star mass and radius is shown with the shaded region on the right-bottom side of the figure by adopting the constraint on $K_\tau$ from experiments at RCNP, i.e., $K_\tau=-550\pm 100$ MeV \cite{Li10}, together with $L=60\pm 20$ MeV as a fiducial value of $L$, which corresponds to $\eta_\tau=59.9-113$ MeV. We also show the allowed region with the empirical relation as a function of $\eta$, assuming the fiducial values of $L=60\pm 20$ MeV and $K_0=240\pm 20$ MeV. For reference, we show the constraints obtained from the astronomical observations, i.e., PSR J0030+0451 and MSP J0740+6620 with NICER, the shaded region with the observation of x-ray bursts, and $1.4M_\odot$ neutron star radius from GW170817 (see text for the details), together with the theoretical mass-radius curves constructed with some EOSs. The shaded region on the left-top side of the figure denotes the excluded region from the causality. 
}
\label{fig:MRconst}
\end{figure}

The mass of a neutron star is one of the most important pieces of information for constraining the EOS. In particular, the massive star is more important to exclude the soft EOS, with which the expected maximum mass does not reach the observed mass. Up to now, the maximum mass precisely determined is $M = 2.08 \pm0.07M_\odot$ (68.3\% credibility) for MSP J0740+6620 \cite{C20,F21}. Owing to the NICER observations, the neutron star mass and radius for PSR J0030+0451 \cite{Riley19,Miller19} and MSP J0740+6620 \cite{Riley21,Miller21} are also constrained. The resultant $68\%$ and $95\%$ confidential levels are shown in Fig. \ref{fig:MRconst}. In addition, the gravitational wave observed from the binary neutron star merger, GW170817, told us the tidal deformability of the neutron star, which leads to the constraint on the $1.4M_\odot$ radius as $R_{1.4}\le 13.6$ km \cite{Annala18}. This constraint must be conservative, while more stringent constraints on the $1.4M_\odot$ radius are also suggested from multimessenger observations and nuclear theory, i.e., $R_{1.4}=11.0^{+0.9}_{-0.6}$ km (90\% confidence) \cite{Capano20} and $R_{1.4}=11.75^{+0.86}_{-0.81}$~km (90\% confidence) \cite{Dietrich20}. Furthermore, the neutron star mass and radius are constrained from the x-ray bursts observations \cite{Steiner13}, although this constraint may strongly depend on the theoretical model. Meanwhile, as a theoretical constraint, the left-top region on Fig. \ref{fig:MRconst} should exclude due to the causality \cite{Lattimer12}. We collect these constraints in Fig. \ref{fig:MRconst}. From this figure, one can see that the constraints from the astronomical observations seem to be still uncertain, but they will gradually become better. In the same figure, we also plot the neutron star mass and radius relations constructed with some of the specific EOSs. We note that the Shen EOS is ruled out from the $1.4M_\odot$ radius constraint, but we show it because the Shen EOS is one of the standard EOSs adopted in many astrophysical studies.

\subsection{Constraints from terrestrial experiments}
\label{sec:MR-Ktau}

Since the neutron star EOS cannot characterize only by the nuclear saturation parameters in a higher density region, the neutron star mass and radius discussed with the nuclear saturation parameters constrained from terrestrial experiments are only for low-mass neutron stars, which are located at the right-bottom side on the neutron star mass and radius relation. Anyway, to connect the neutron star mass and radius with the nuclear saturation parameters, the empirical relations for neutron star mass and radius (or gravitational redshift) must be useful. With the empirical relation for the neutron star mass and gravitational redshift as a function of $\eta$~\cite{SIOO14}, we have already discussed the expected region in the neutron star mass and radius relation \cite{SNN22}, e.g., the allowed region with the fiducial values of $L$ and $K_0$, i.e., $L=60\pm 20$ and $K_0=240\pm 20$ MeV, is shown as in Fig. \ref{fig:MRconst}, where the central density is considered up to twice the saturation density. In a similar way, with the empirical relations derived in this study as a function of $\eta_\tau$, the expected region in the neutron star mass and radius relation can be shown with the shaded region in Fig. \ref{fig:MRconst}, assuming that $L=60\pm 20$ and $K_\tau=-550\pm 100$ MeV, where the central density is considered up to threefold the saturation density and the stellar models with twice the saturation density are also denoted with the dotted line in the shaded region.

\begin{figure}[tbp]
\begin{center}
\includegraphics[scale=0.6]{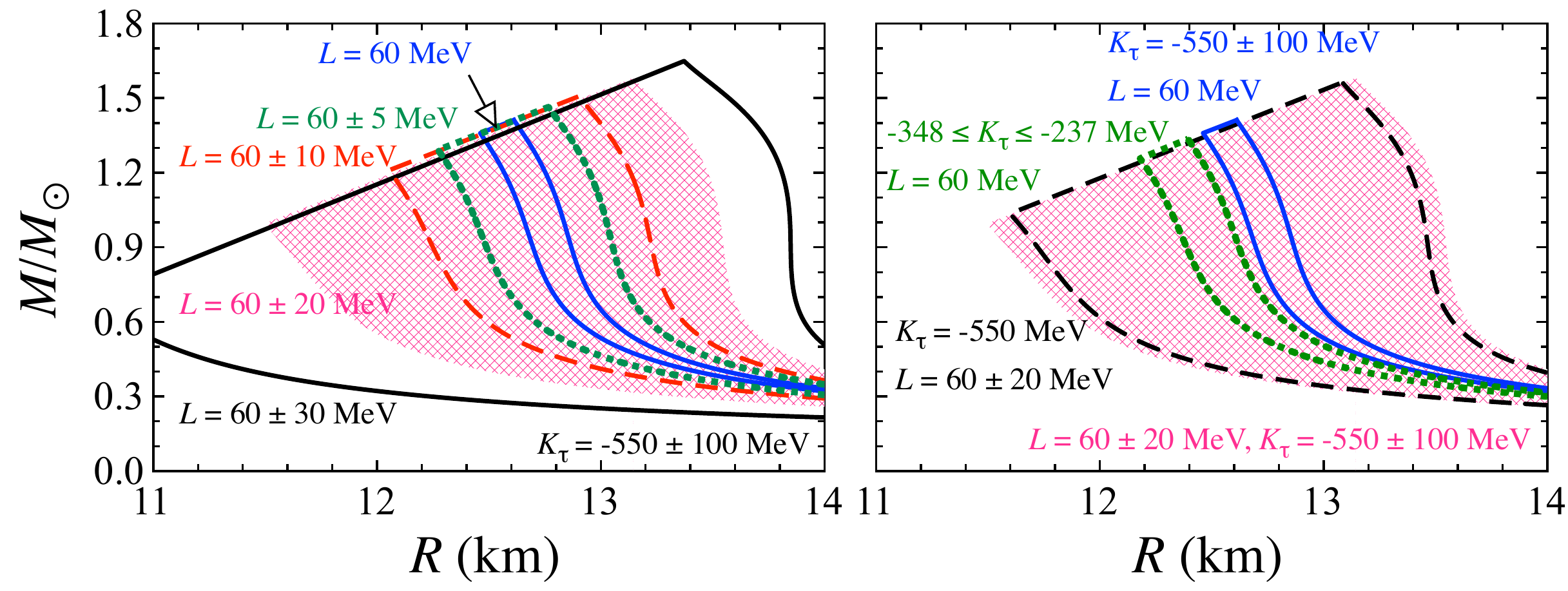}
\end{center}
\caption{
The expected region of the neutron star mass and radius, using the empirical relations as a function of $\eta_\tau$. The shaded region corresponds to the constraint on neutron star mass and radius with $L=60\pm 20$ and $K_\tau=-550\pm 100$ MeV, which is the same as the shaded region shown in Fig. \ref{fig:MRconst}. In the left panel, the expected regions are additionally shown for $L=60\pm 30$, $L=60\pm 10$, $L=60\pm 5$, and $L=60$ MeV  together with $K\tau = -550\pm 100$ MeV.
In the right panel, the region enclosed with the dashed, solid, and dotted lines correspond to the region expected with $L=60\pm 20$ and $K_\tau=-550$ MeV, $L=60$ and $K_\tau=-550\pm 100$ MeV, and $L=60$ and $-348\le K_\tau \le -237$ MeV respectively. 
}
\label{fig:MRKtau}
\end{figure}

Finally, in Fig. \ref{fig:MRKtau}, we show the expected region of the neutron star mass and radius, adopting the specific value of $L$ and $K_\tau$, e.g., the shaded region for $L=60\pm 20$ and $K_\tau = -550\pm 100$ MeV as in Fig. \ref{fig:MRconst}. In addition, in the left panel, the expected regions are shown for $L=60\pm 30$, $L=60\pm 10$, $L=60\pm 5$, and $L=60$ MeV with $K_\tau = -550\pm 100$ MeV, while in the right panel, the region enclosed with the dashed line, the solid line, and the dotted line correspond to the expected mass and radius for $L=60\pm 20$ and $K_\tau = -550$ MeV; $L=60$ and $K_\tau = -550\pm 100$ MeV; and $L=60$ and $-348 \le K_\tau \le -237$ MeV, respectively. From this figure, one can obviously see that the $L$ dependence is much stronger than the $K_\tau$ dependence for determining the neutron star properties, but still $K_\tau$ also plays an important role in neutron star mass and radius relation.

\section{Conclusion}
\label{sec:Conclusion}

To discuss the neutron star mass and radius with the constraints on the nuclear saturation parameters obtained from terrestrial experiments, the empirical relations for the neutron star mass and radius (or gravitational redshift) as a function of nuclear saturation parameters must be useful, if they exist. In this study, we find a suitable combination, $\eta_\tau$, of the nuclear saturation parameters for expressing neutron star mass and gravitational redshift and derive their empirical relations, focusing on the saturation parameters for asymmetric nuclear matter. With the resultant empirical relations, which are applicable up to threefold the saturation density of symmetric nuclear matter, one can evaluate the neutron star mass and gravitational redshift within $\sim 10\%$ accuracy, and the radius within a few \% accuracies. In addition, with these empirical relations, we discuss the expected region in the neutron star mass and radius relation, assuming the experimental values of saturation parameters, together with the constraints from the astronomical observations. Furthermore, we find a tight correlation between $\eta_\tau$ and $\eta$, which is another suitable combination of saturation parameters for symmetric nuclear matter. With this correlation, we derive the constraint on the isospin dependence of incompressibility for asymmetric nuclear matter, $K_\tau$, as $-348\le K_\tau\le -237$ MeV, assuming the fiducial value of the density-dependent nuclear symmetry energy and the incompressibility for symmetric nuclear matter.
The value of $K_\tau$ derived in this study shows the discrepancy from the values deduced from experiments. This may be because the experiments have been done using finite nuclei. As pointed out in Ref. \cite{Stone14}, the surface term or the mass number dependence especially in the expansion of the incompressibility in finite nuclei should be evaluated in such a way as to appropriately deduce the incompressibility and its isospin dependence of the nuclear matter. Systematic measurement of the incompressibility in finite nuclei in a wide mass-number range may provide an opportunity to evaluate such a finite system effect. 
In addition, we found $\eta$ and $\eta_\tau$ through trial and error. Since these are good quantities to express the low-mass neutron stars and we also find a strong correlation between $\eta$ and $\eta_\tau$, there may exist a physical meaning behind these quantities. If one will find such a physical meaning, it must tell us a new aspect of nuclear physics, which has never been known up to now.

\acknowledgments
We are grateful to Hajime Togashi for preparing the EOS data adopted in this study and also to David Blaschke and Tomoya Naito for valuable comments. This work is supported in part by Japan Society for the Promotion of Science (JSPS) KAKENHI Grant Numbers 
JP19KK0354,  
and
JP21H01088,  
and by Pioneering Program of RIKEN for Evolution of Matter in the Universe (r-EMU).



\end{document}